\documentclass[aps,twocolumn,showpacs,amsmath,floatfix]{revtex4}
\usepackage{graphicx}%
\usepackage{dcolumn}
\usepackage{color}
\usepackage{amsmath}
\usepackage{here}

\begin{document}
\newcommand{\be}{\begin{equation}}
\newcommand{\ee}{\end{equation}}
\newcommand{\rojo}[1]{\textcolor{red}{#1}}

\title{Generation of spin-wave dark solitons with phase engineering}

\author{Bengt Bischof$^1$, Andrei N. Slavin$^{1,2}$, Hartmut Benner$^1$, and Yuri Kivshar$^{3}$}

\affiliation{$^1$Institut f\"ur Festk\"orperphysik, Technische
Universit\"at Darmstadt, D-64289 Darmstadt, Germany\\
$^2$Department of Physics, Oakland University, Rochester, Michigan
48309\\ $^3$Nonlinear Physics Centre, Research School of Physical
Sciences and Engineering, Australian National University,
Canberra, ACT 0200, Australia}

\begin{abstract}
We generate experimentally spin-wave envelope dark solitons from
rectangular high-frequency dark input pulses with {\em externally
introduced} phase shifts in yttrium-iron garnet magnetic films. We
observe the generation of both {\em odd} and {\em even} numbers of
magnetic dark solitons when the external phase shift varies. The
experimental results are in a good qualitative agreement with the
theory of the dark-soliton generation in magnetic films developed
earlier [Phys. Rev. Lett. {\bf 82}, 2583 (1999)].
\end{abstract}


\maketitle

\section{Introduction}

Dark solitons have been predicted theoretically and observed
experimentally in different types of dispersive or diffractive
nonlinear media, including optical and magnetic
systems~\cite{book}. Recently, dark solitons have been generated
in Bose-Einstein condensates of dilute atomic gases~\cite{BEC} by
engineering the phase of the macroscopic wave function with a
technique known as {\em phase imprinting}, earlier developed to
create optical and matter-wave vortices~\cite{book}. Phase
imprinting is a new tool of manipulating coherent matter waves,
and it is described as shining an off-resonance laser on a
Bose-Einstein condensate in order to create phase steps between
its different parts. A given phase step defines the parameters of
dark solitons travelling in the same or opposite directions, and
the total number of the generated matter-wave dark solitons which
can be {\em either odd or even}~\cite{BEC2}.

As a matter of fact, these recent results on the generation of
dark solitons in Bose-Einstein condensates can be compared with
and linked to much earlier experimental studies of microwave
magnetic-envelope spin-wave solitons in yttrium-iron garnet (YIG)
magnetic films~\cite{first}. Indeed, the first observation of {\em
dark solitons} in YIG magnetic films~\cite{Patton} revealed
unusual features of the dark-soliton generation observed as a
change of the total number of generated spin-wave dark solitons
from even to odd with the growth of the input power. These
observations have later been explained theoretically by employing
the concept of the so-called {\em induced spatial phase
shift}~\cite{SKB}, which is closely related to the concept of
phase imprinting, as was also discussed in Ref.~\cite{BEC2}.
Recently, a further attempt to employ the phase manipulation
technique for generating single and multiple dark magnetic
solitons has been undertaken in Ref.~\cite{Serga}.

The purpose of this paper is twofold.  First, we extend the
concept of phase imprinting implemented for dark solitons and
vortices in Bose-Einstein condensates to the field of magnetic
solitons, and generate experimentally single and multiple
spin-wave envelope magnetic dark solitons from rectangular dark
input high-frequency pulses with {\em externally introduced phase
shifts} in YIG magnetic films. Second, we provide a direct
verification of the theory developed earlier in Ref.~\cite{SKB}.

The paper is organized as follows. In Section~\ref{model} we
introduce our model which was first suggested in the pioneering
papers~\cite{first} and is described by the cubic nonlinear
Schr\"odinger (NLS) equation for the magnetic field envelope.
Section~\ref{theory} summarizes the basic theoretical results for
the generation of dark solitons by an input pulse with a jump
across the low-intensity region~\cite{SKB}. The main
Sec.~\ref{exp} presents our experimental results which are shown
to be in a good agreement with the basic theoretical predictions.

\section{Model}
\label{model}

We consider the evolution of a spin wavepacket in the form $\Psi
(x,t) =u(x,t)\exp \{i(k_0x-\omega_0 t)\}$, where the slowly
varying complex envelope $u(t,x)$ is described by the
dimensionless nonlinear Schr\"odinger (NLS) equation~\cite{first},
\begin{equation}
i\left(\frac{\partial u}{\partial t}+v_g\frac{\partial u}{\partial
x}\right)+ \frac{1}{2}D\frac{\partial^2 u}{\partial
x^2}-N|u|^2u=0,
 \label{eq:nls}
\end{equation}
where $v_g=\partial \omega/\partial k$ is the group velocity
evaluated from the spin-wave nonlinear dispersion $\omega(k,
|u|^2)$ at the carrier wavenumber $k_0=k(\omega_0)$, $D=\partial^2
\omega/\partial k^2$ is the coefficient of linear dispersion, and
$N=\partial \omega/\partial |u|^2$ is the coefficient of
nonlinearity. In the derivation of Eq.~(\ref{eq:nls}) dissipation
is neglected~\cite{first}.

In the case $DN>0$, Eq.~(\ref{eq:nls})  has a solution in the form
of a {\it dark soliton} that can be written as
follows~\cite{HasTap}:
\begin{equation}
u(x,t)=u_0 \tanh \xi \, \exp\{i(K x-\Omega t)\},
\label{eq:dark_soliton}
\end{equation}
where  $\xi=(t-t_0-x/v_s)/ \tau_0$. Solution
(\ref{eq:dark_soliton}) describes a localized dip in the
continuous wave (CW) background with the half-width $\tau_0$, the
center $t_0$, and the velocity $v_s=v_g+DK$. In the phase factor
of the dark solitons, $K$ is the soliton wave number and
$\Omega=v_gK+ (1/2)DK^2+N|u_0|^2$ is the soliton frequency shift,
which are the nonlinearity-induced corrections to the wavenumber
$k_0$ and carrier frequency $\omega_0$.

Solution~(\ref{eq:dark_soliton}) is a special case of a dark
soliton with a modulation factor $A=1$, i.e. when the dark soliton
has a minimum amplitude of $u=0$ (the so-called ``black soliton").
Due to the $\tanh$-function, this soliton has an overall phase
shift of $\Phi=\pi$. However, such pulses usually do not appear in
experimental situations. Instead, a more general form of a dark
soliton~\cite{HasTap} should be used,
\begin{equation}
u(x,t)=u_0(1-A^2{\rm sech}^2\xi)^{1/2}
\exp\{i[\tilde{K}x-\tilde{\Omega}t+\sigma(\xi)]\},
\label{eq:grey_soliton_solution}
\end{equation}
where the phase factor has the form:
\begin{eqnarray}
\sigma(\xi)=\sin^{-1}\left(\frac{A\tanh\xi}{\sqrt{1-A^2{\rm
sech}^2\xi}}\right),
\label{eq:phase_factor} \\
\nonumber \tilde{K}=K-\frac{\sqrt{1-A^2}}{v_s\tau_0A}, \quad
\tilde{\Omega}=\Omega-\frac{\sqrt{1-A^2}}{\tau_0A}.
\end{eqnarray}
Solution~(\ref{eq:grey_soliton_solution}) describes a single
``grey soliton'' with an arbitrary value of the modulation depth
$A$. For $A=\pm 1$, Eq.~(\ref{eq:grey_soliton_solution})
transforms into the "black" soliton~(\ref{eq:dark_soliton}).

The important condition for such dark solitons to exist is the
phase shift given by Eq.~(\ref{eq:phase_factor}) that has to be
present in the carrier wave~\cite{book}. In the experimental
situation, there is no such total phase shift and, therefore, only
even-numbered symmetric pairs of dark solitons with equal
modulation $|A|<1$ can be generated. In these pairs, the phase
shifts of the individual dark solitons have opposite signs and
they compensate one another, so the total phase shift adds up to
vanish, $\phi=0$. Under such conditions, excitation of odd numbers
of dark solitons seems, in general, impossible~\cite{book}.

\section{Theoretical background}
\label{theory}

\begin{figure}
\centerline{\includegraphics[width=3.0in]{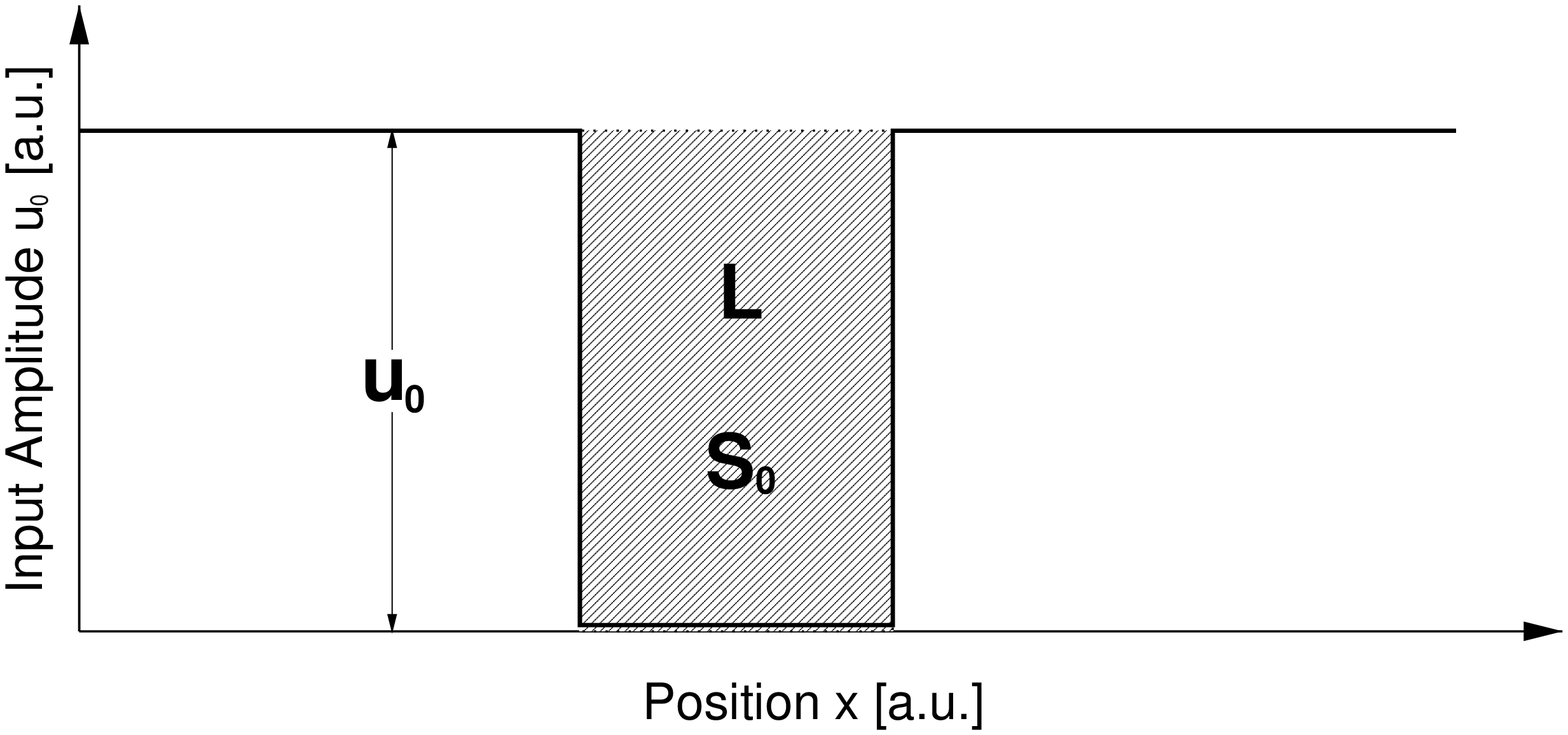}}
\centerline{\includegraphics[width=3.0in]{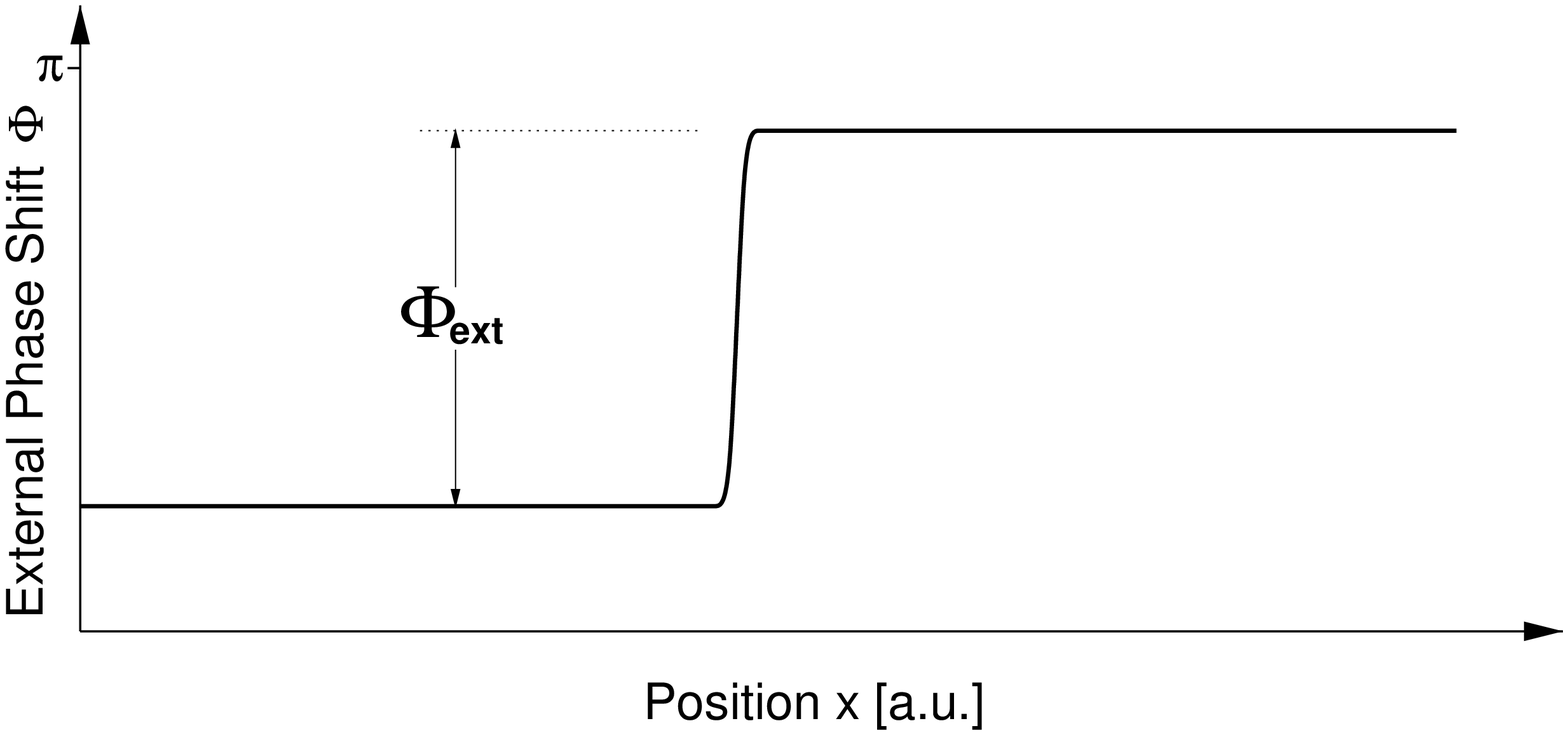}}
\caption{Top: Dark input pulse with amplitude $u_0$, length $L$,
and area $S_0$ (shaded). Bottom: Corresponding variation of the
externally introduced input phase (not to
scale)~\cite{phasensprungdefinition}.} \label{fig:input}
\end{figure}

Recently, it was shown analytically and numerically that when an
input pulse {\em without} any initial phase modulation enters a
nonlinear dispersive medium, the generated localized wave acquires
an {\em induced spatial phase shift} accumulated during its
generation~\cite{SKB}. Such a phase shift is negligible for large
group velocities, e.g., for optical solitons in fibers, but it
becomes important for spin waves in magnetic films. Moreover, if
the initial phase shift (Fig.~\ref{fig:input}) is included in the
inverse scattering transform~\cite{zakharov}, the results can be
employed to explain~\cite{SKB} the specific features of the
generation of both odd and even numbers of dark solitons observed
in experiment~\cite{Patton}.

In order to demonstrate the role of the initial phase shift, we
summarize the results for the dark-soliton generation in the
framework of the normalized NLS equation,
\begin{equation}
i\frac{\partial u}{\partial t}+\frac{\partial^2 u}{\partial
x^2}-n|u|^2u=0 \label{eq:rnls}
\end{equation}
Equation~(\ref{eq:rnls}) follows from Eq.~(\ref{eq:nls}) after
rescaling, by assuming the reference frame moving with the group
velocity $v_g$, and renormalizing the variables $t$ and $x$. The
input pulse is characterized by the phase shift $\phi$ and the
dimensionless value $S_0$, which is defined as
$S_0=|u_0|L\sqrt{(N/D)}$ and is usually called the input pulse
{\it area} as it is proportional to the carrier wave amplitude
$u_0$ multiplied by (spatial) pulse length $L=v_gT$. The soliton
wavenumber shift can then be presented as $K=2\pi/L$.

According to Ref.~\cite{SKB}, the number and symmetry of the
excited sequence of dark solitons are defined by the
transcendental equation for the soliton eigenvalue $\nu$,
\begin{equation}
\tan(\nu-\phi)=\frac{\sqrt{S_0^2-\nu^2}}{\nu}.
\label{eq:solifeld}
\end{equation}
Every solution of this equation for the real eigenvalue $\nu_n$
corresponds to a dark soliton with the modulation depth
$\kappa_n=\sqrt{S_0^2-\nu_n^2}/L$, where $\kappa_n$ has the
dimension of the renormalized amplitude, $|u_0|\sqrt{(N/D)}$. The
right-hand side of Eq.~(\ref{eq:solifeld}) is always
point-symmetric around the origin whereas for the $\tan$-function
this is only the case when $\phi=\Phi/2$ is
zero~\cite{phasendefinition}. In this latter case,
Eq.~(\ref{eq:solifeld}) yields pairs of eigenvalues with the same
absolute value but with opposite signs. This explains why only
even-numbered symmetric pairs of dark solitons are predicted by
the earlier theory~\cite{zakharov}. Note that a dark soliton with
zero amplitude minimum is defined by $\kappa_n=|u_0|(N/D)^{1/2}$,
i.e. at $\nu_n=0$. This is only the case for $\phi=\pi/2$ and odd
multiples thereof.

\begin{figure}
\centerline{\includegraphics[width=3.0in]{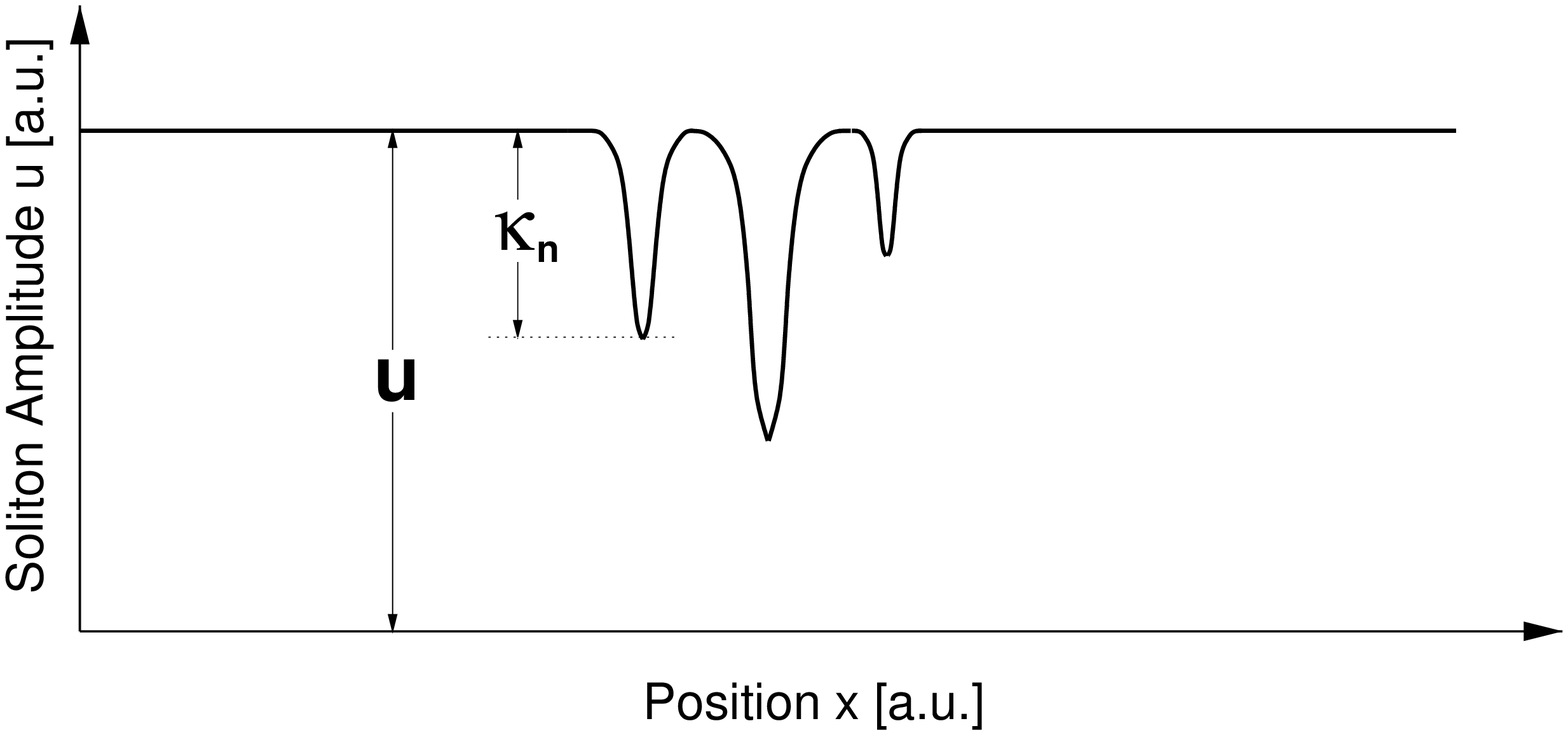}}
\centerline{\includegraphics[width=3.20in]{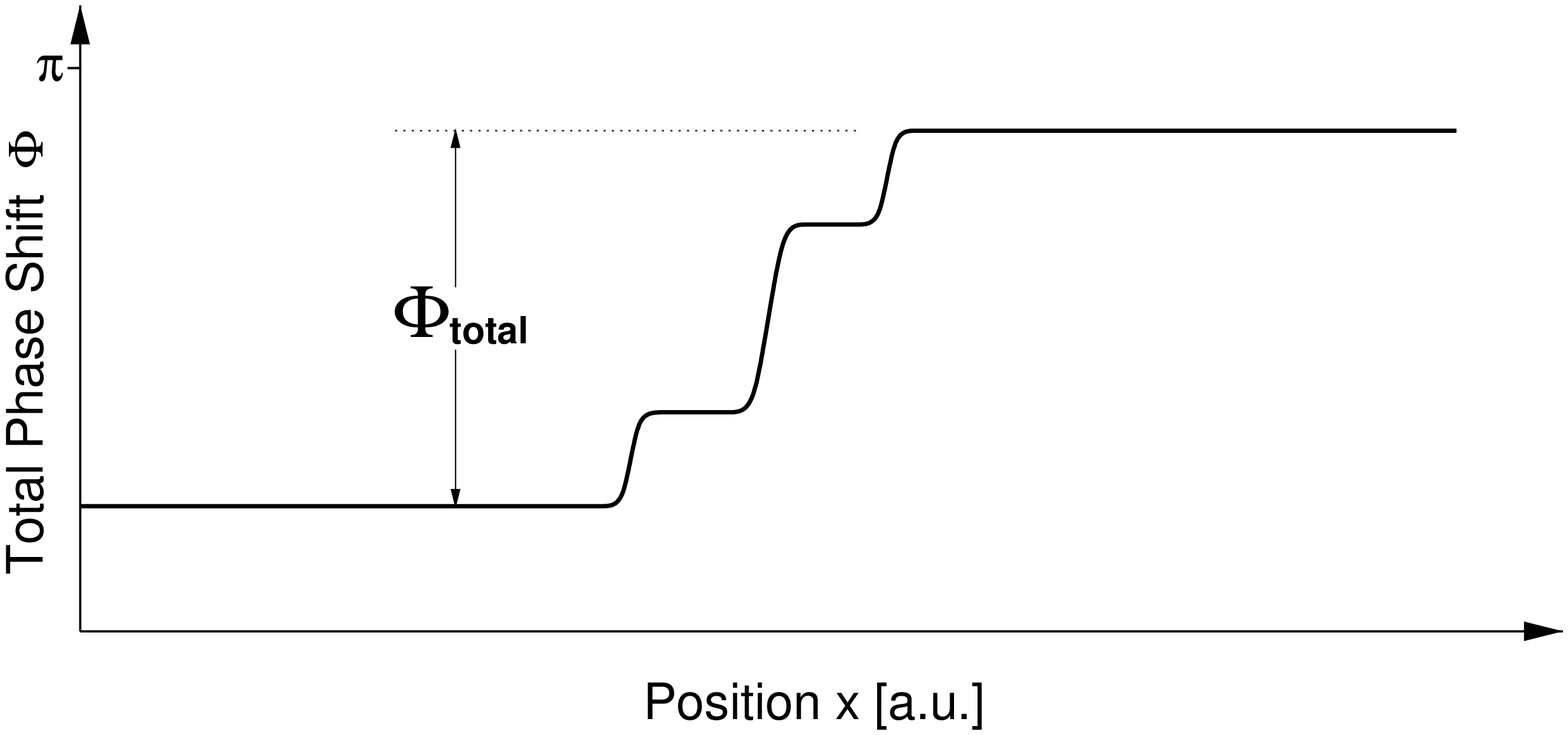}}
\caption{Top: Schematic for the generation of $N$=3 dark solitons.
Bottom: Corresponding simplified phase structure.}
\label{fig:3sol}
\end{figure}

\begin{figure}
\centerline{\includegraphics[width=3.1in]{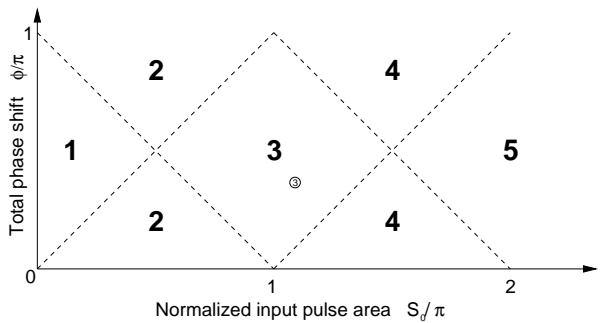}}
\caption{Number of dark solitons in the plane $(S_0, \phi)$. The
circle indicates the $N$=3-soliton shown in Fig.~\ref{fig:3sol}.
The separation lines between different numbers of solitons are at
$\phi=M\pi\pm S_0$, $M=0,\pm1,\ldots$, the lines of exact symmetry
are horizontal lines at $\phi=N\pi/2$. For even $N$, there will be
pairs of symmetric solitons; for uneven $N$, a single soliton of
100\% modulation depth surrounded by pairs of symmetric solitons.}
\label{fig:solifeld}
\end{figure}

A finite initial value of the phase shift $\phi \neq 0$ allows to
predict the generation of asymmetric or symmetric, odd- or
even-numbered dark soliton pulses. The typical structure of an
$N$=3-soliton state and the corresponding variation of its phase
are schematically shown in Fig.~\ref{fig:3sol}. Such a soliton
sequence is characterized by the overall number of dark solitons
and the modulation of each individual dark soliton. According to
Eq.~(\ref{eq:solifeld}), these values are completely determined by
the input pulse area $S_0$ and the total phase shift $\phi$. This
suggests~\cite{SKB} the use of the parameter plane $(S_0,\phi)$ as
presented in Fig.~\ref{fig:solifeld}. Changing the pulse length
and/or input amplitude is equivalent to moving on the horizontal
axis. A change in the total phase shift is denoted by a shift on
the vertical axis. Every point in that parameter plane can be
assigned a value of $S_0$ and $\phi$, therefore it is directly
connected to a dark soliton of distinct eigenvalues $\nu_n$.

In Ref.~\cite{SKB}, it was shown that in physical systems with low
values of the group velocity, an additional phase shift is
acquired during the process of pulse excitation. However, in the
general case, {\em the total phase shift} in the input pulse
carrier wave may originate either from the pulse forming setup,
i.e. it is provided externally, or from an intrinsic mechanism in
the physical system. In this paper, we present a systematic study
of the formation of magnetic dark solitons with a variable
external phase shift $\phi_{\rm ext}$. Using these data, we
verify, in particular, whether an acquired phase shift $\phi_{\rm
acq}$ is present at all and whether it fits the theoretically
expected values.

In essence, the results of Ref.~\cite{SKB} predict that an
additional phase shift $\phi_{\rm acq}$ acquired during the
process of dark-soliton generation can be written in the form,
\begin{equation}
\label{eq:phiint} \phi_{\rm
acq}=\frac{bT}{T_D}\left[1+\frac{1}{2}\left(\frac{S_0}{\pi}\right)^2\right],
\end{equation}
where $T_D=(v_g^2 T^2/\pi^2|D|)$ is the (positive) dispersion
time. The phenomenological factor $b$ (introduced in the
theory~\cite{SKB} and evaluated there to be close to 2) takes into
account the influence of the finite length of the background CW
signal.

Usually, the sign of $\phi_{\rm acq}$ would not be important due
to the symmetry of Eq.~(\ref{eq:solifeld}). However, as we are
using a phase shifter in the experiment to produce the external
phase $\phi_{\rm ext}$,  the relative sign of these two phase
shifts is important. In the experiments described below, the sign
of the acquired phase shift  $\phi_{\rm acq}$ was opposite to that
of the external phase shift $\phi_{\rm ext}$. Thus, the total
phase shift of the input pulse is expressed as
\begin{equation}
\label{eq:phitot} \phi_{\rm total}=\phi_{\rm ext}-\phi_{\rm acq},
\end{equation}
where $\phi_{\rm acq}$ is determined by Eq.~(\ref{eq:phiint}).
Note that there is always a fixed `offset` phase shift, even in
the limit $S_0=0$. This means that all dark solitons in magnetic
systems acquire a phase shift during the process of their
generation. However, if the group velocity and, therefore, $T_D$
is large, as we have for the case of dark solitons in
optics~\cite{book}, the phase $\phi_{\rm acq}$ is small and will
remain unimportant. In this case, the dark solitons are always
placed on a horizontal line at $\phi=0$ on the $(S_0,\phi)$-plane
which is equivalent to the production of even-numbered symmetric
pairs of dark solitons.

If the input pulse parameters are in a range that $T_D\approx T$,
the acquired phase shift may manifest itself by producing
asymmetric and/or odd-numbered dark solitons, and the points
corresponding to these solitons on the parameter plane of
Fig.~\ref{fig:solifeld}  will belong to the parabolic curve
defined by Eq.~(\ref{eq:phiint}) and having a minimum at the point
$S_0$=0. This case was realized in the pioneering experiments
performed by Patton's group~\cite{Patton}.

On the other hand, if the constant and positive external phase
shift $\phi_{\rm ext}>0$ is introduced externally, while duration
$T$ of the input pulse is kept constant and the input wave
amplitude (and, therefore, $S_0$) is increased, the dependence
$\phi_{\rm total}(S_0)$  has the form of a quadratic parabola with
a maximum at $S_0$=0 [see Eq.~(\ref{eq:phitot})]. As will be seen
below, it is this case that is realized in our current
experiments. Therefore, the points corresponding to different
output solitonic pulses obtained in our experiments with a
positive external phase shift and increasing input wave amplitude
belong to parabolas having a maximum at $S_0$=0 (see
Fig.~\ref{fig:solifeldtheo15} below).

\section{Experimental results}
\label{exp}

To study the effects produced by an external phase shift on the
generation of magnetic dark solitons, we need to design an
experimental setup with the following characteristics: (i) the
setup should include a nonlinear dispersive waveguide capable of
supporting spin-wave dark solitons, i.e. the necessary condition
for the formation of dark solitons, $N D>0$, should be fulfilled;
(ii) the parameters of the pulses propagating in the waveguide
should fit the range where the induced phase shift defined by the
Eq.~(\ref{eq:phiint}) has a large enough value and could be
detected experimentally; (iii) there should be a  technical
possibility to introduce {\em an external phase shift} between the
leading and trailing fronts of the input pulse, similar to the
phase shift introduced by the method of phase imprinting in
Bose-Einstein condensates or nonlinear optics~\cite{book}.

For our experiments, we have chosen magnetostatic spin waves
propagating in a thin quasi-one-dimensional waveguide made of a
single-crystal YIG film.  Nonlinear properties of these waves are
well understood, and the formation of both dark and bright
solitons has been observed~\cite{first,Patton}. Moreover, the
measurement techniques incorporating the so-called {\it delay
line} setup are well established. In our case, it is especially
important that one can easily introduce an external phase shift in
the input pulse simply using a microwave phase shifter. The most
important feature of the magnetostatic waves in the context of
this study is their comparatively {\em low group velocity} that
should result in a substantial value of the induced phase shift
defined by Eq. (\ref{eq:phiint})(see Ref.~\cite{SKB} for details).

The experiments are performed on a 1.5 mm wide and 7$\mu$m thick
YIG film waveguide which, due to its relatively small width, could
be considered as quasi one-dimensional. The film waveguide is
fixed in a standard delay line setup consisting of two strip-line
antennas of 45$\mu$m width and 9.2 mm separation. A microwave
frequency of 5.065 GHz is used. A magnetic bias field of 1107.5 Oe
strength is applied tangentially to the film surface and
perpendicularly to the propagation direction of spin waves and,
therefore, the conditions for the excitations of magnetostatic
surface waves are fulfilled. The value of the magnetic field is
chosen so as to place the working point frequency in the middle of
the spin-wave spectrum, letting all Fourier components of the
dark-pulse signal pass through.

\begin{figure}
\centerline{\includegraphics[width=3.1in]{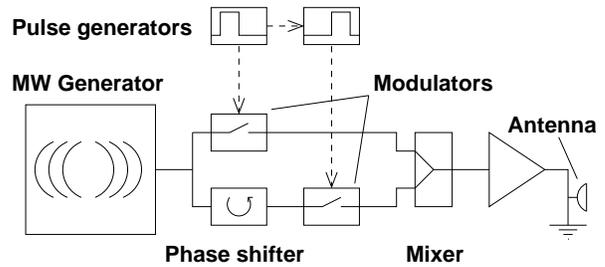}}
\caption{Schematic diagram of the experimental setup for
generating dark pulses. The two pulse generators trigger the two
microwave modulators with a fixed delay. Variation of the length
of the first pulse provides the dark-pulse duration. The phase
shifter introduces the external phase shift $\phi_{\rm ext}$.}
\label{fig:pulsformung}
\end{figure}

A variable phase shift in the input pulse is created in the
following way: a continuous microwave signal produced by a
sweep-generator is divided into two parts, which are then
individually controlled by microwave modulators connected to the
pulse generators. A standard microwave phase shifter is introduced
in one of the modulation channels. Then, the two resulting
pulse-modulated signals are combined together in a microwave mixer
device, and are supplied to the input antenna of the experimental
delay line (see Fig.~\ref{fig:pulsformung}).

\begin{figure}
\centerline{\includegraphics[width=2.8in]{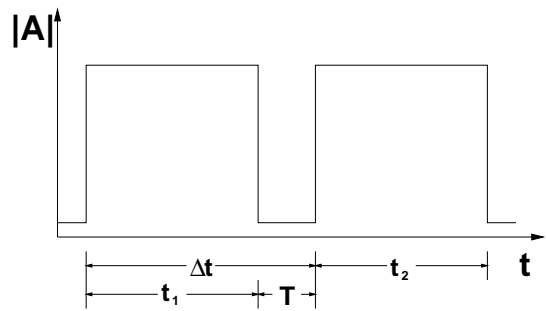}}
\caption{Timing scheme of the dark-pulse generation setup; $t_1$
is varied, thus changing the value of the parameter $T$.}
\label{fig:zeitschema}
\end{figure}

In the experimental scheme described above, the trailing edge of
the bright rectangular pulse created in the first modulation
channel provides the leading edge of the input dark pulse, while
the leading edge of the bright rectangular pulse created in the
second modulation channel provides the trailing edge of the input
dark pulse. The phase shift $\phi_{\rm ext}$ is introduced by the
variable phase shifter inserted in the second modulation channel.
Thus, the duration $T$ of the formed dark input pulse (see
Fig.~\ref{fig:zeitschema}) is determined by the difference of the
time delay $\Delta t$ between the leading edges of both bright
rectangular pulses and the duration $t_1$ of the bright pulse in
the first modulation channel $T=\Delta t-t_1$. The duration $T$ of
the input dark pulses in our experiment could be changed in 1 ns
steps, and the duration of the CW background before and after the
input dark pulse is controlled by changing the durations $t_1$ and
$t_2$ in the first and second modulation channels, respectively.
In our experiments, the values of $t_1$ and $t_2$ are chosen to be
constant and equal, $t_1$ = $t_2$ = 1$\mu$s. The input pulse
sequences of the form shown in Fig.~\ref{fig:zeitschema} are
produced with a repetition rate of about 100 kHz.

The magnitude of the external phase shift introduced into the dark
input pulse is controlled by the variable phase shifter in the
second modulation channel. The zero point and the scale
calibration of the phase shifter are established by adjusting it
to the points of minimum and maximum interference of two
overlapping pulses. The sign of the phase shift is established by
calibrating it at different frequencies.

The signal is recorded with a digital oscilloscope and a transient
recorder (500 MHz) through a diode detector. The signal received
by the output antenna is amplified using a low-noise small-signal
amplifier to drive the detector diode in its optimum sensitivity
range and to achieve the highest signal-to-noise ratio possible.
The output voltage in this range is proportional to the power at
the input of the detector diode.

In our experiment the variable parameters are: (i) input power
$P_{\rm in}$, (ii) duration of the dark input pulse $T$, and (iii)
external phase shift $\phi_{\rm ext}$. The input power is varied
in the interval $P_{\rm in}=1.6 \, {\ldots} \, 100$ mW, while the
dark pulse duration is varied in the interval $T=10 \, {\ldots}\,
30$ ns  in steps of 5 ns.  We use the following values of the
external phase shift: 0.17$\pi$, 0.42$\pi$, 0.6$\pi$, 0.72$\pi$,
and 0.95$\pi$.

The experimental results are summarized using the parameter plane
of Fig.~\ref{fig:solifeld}. The profiles of the output pulses
corresponding to the fixed values of the input power $P_{\rm in}$,
dark input pulse duration $T$, and external phase shift $\phi_{\rm
ext}$ are recorded, and the corresponding points are placed on the
parameter plane Fig.~\ref{fig:solifeld}. Counting the number of
the minima in the output pulse profile corresponding to a
particular set of input parameters,  we can determine the number
of solitons and compare it with the theory.

However, to compare the experimental results with the theory, the
dispersion and nonlinearity parameters of the magnetostatic waves
propagating in the YIG film wave\-guide should be either measured
or calculated. First, the group velocity at the working point is
determined by measuring the time of the pulse propagation between
the input and output antennae of the setup. This velocity is found
to be equal to $5.5\cdot10^6$cm/s ($\pm2\%$) which is close to the
theoretical estimate of the group velocity of $5.3\cdot10^6$cm/s
($\pm2\%$) obtained from Eq.~(55) in Ref.~\cite{Wigen}.

The value of dispersion $D$ was established by comparing $v_g$ at
several frequencies $\pm50$MHz above and below the carrier
frequency $\omega_0$. Using a polynomial fit function of
$v_g(\omega)$, the value of $D$ at the working point was
determined to be $-7.2\cdot10^3$cm$^2/$s ($\pm20\%$).  The large
error is due to the small differences of $v_g$ at small frequency
intervals. The theoretical estimation of the dispersion
coefficient $D$ done using Eq.~(55) from Ref.~\cite{Wigen} gives
the value of $-7.1\cdot10^3cm^2/s\pm20\%$. Thus, in the
calculations below we use the values: $v_g=5.5\cdot10^6$cm/s
($\pm2\%$) and $D=-7.2\cdot10^3$cm$^2/$s ($\pm20\%$).

The value of the nonlinearity coefficient $N$ could not be
measured directly. Thus, it was calculated using Eq. (52) from
Ref.~\cite{Wigen}  to give  $N=9.5\cdot 10^9 s^{-1}$.

\begin{figure}
\centerline{\includegraphics[width=3.3in]{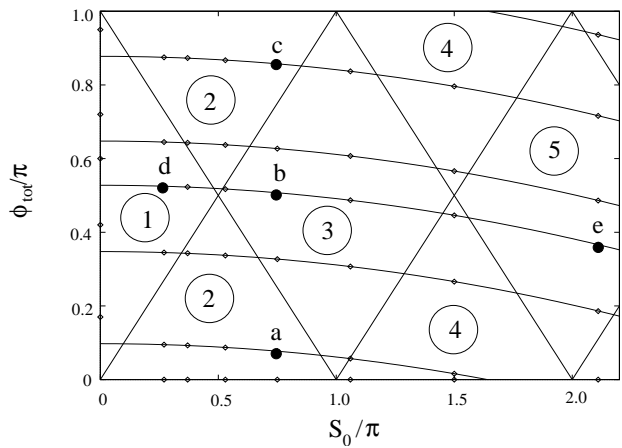}}
\caption{Experimental results for the parameters of dark solitons
with $T=20ns$ on the plane of $\phi_{\rm total}$ vs. $S_0$. The
marked points correspond to the soliton profiles shown in
Fig.~\ref{fig:oscillograms}. } \label{fig:solifeldtheo15}
\end{figure}

The results of our experiments for the pulse duration of $T$ = 20
ns are presented in Fig.~\ref{fig:solifeldtheo15}. Two adjustable
parameters are used to place the experimental points on this
graph.  The first parameter is the coefficient relating the input
power $P_{\rm in}$ to the normalized amplitude of the spin wave
$u_0$, i.e. $|u_0|^2$ = $BP_{\rm in}$. It defines the relation
between the input power $P_{\rm in}$ and the parameter $S_0$ used
in the theoretical formula~(\ref{eq:phiint}) and, therefore, it
determines the scale of the graph along the horizontal axis.
Similar to Ref.~\cite{SKB}, we chose $B$ assuming that at the
input power of $P_{\rm in}= 175$mW, when thermal effects begin to
manifest themselves, we have a typical value of the spin wave
amplitude in the film equal to $|u_0|$ = 0.07 (see
Ref.~\cite{Dudko}). For the conditions of our experiment this
yields $B= 27\cdot 10^{-3} W^{-1}$, which is about three times
larger than the similar value in Ref.~\cite{Patton}. The
difference may be attributed to a better matching between the
impedances of the supply line and the input micro-strip antenna.

The second adjustable parameter is the phenomenological
coefficient $b$ in Eq.~(\ref{eq:phiint}) which describes the
effect of a finite CW background of the input dark pulse and
determines the scale of the graph in Fig.~\ref{fig:solifeldtheo15}
along the vertical axis. Similar to Ref.~\cite{SKB}, we take
$b=2$.

The parabolas presented in Fig.~\ref{fig:solifeldtheo15} are
computed using Eq.~(\ref{eq:phitot}), and they correspond to
different values of the external phase shift 0.17$\pi$, 0.42$\pi$,
0.60$\pi$, 0.72$\pi$, and 0.95$\pi$, so that the lowest parabola
corresponds to the smallest external phase shift of 0.17$\pi$. The
values of the initial (linear) phase shifts corresponding to the
initial point $S_0 = 0$ are computed as
\begin{equation}
\label{eq:phitot_0} (\phi_{\rm total})_0=\phi_{\rm ext}-2T/T_D.
\end{equation}

\begin{figure}
\centerline{\includegraphics[width=3.0in]{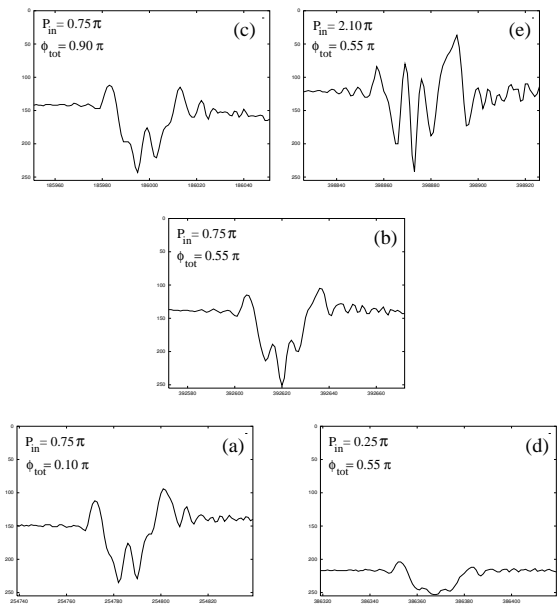}}
\caption{Example of a series of measured dark solitons with equal
pulse duration of 20 ns. Left row: equal input power $P_{\rm
in}=-9$dBm (equivalent to $S_0=0.75\pi$), but different external
phase shifts $\phi_{\rm ext}=0.17\pi$(a), 0.60$\pi$(b) and
0.95$\pi$(c). Right row: equal external phase shift $\phi_{\rm
ext}=0.60\pi$, but different input power $P_{\rm in}=-18$dBm (d),
-9dBm (b) and 0dBm (e).  The soliton profiles correspond to the
vertically and horizontally aligned points in
Fig.~\ref{fig:solifeldtheo15}.} \label{fig:oscillograms}
\end{figure}

The points on the parabolas correspond to the points where the
experimental oscillograms of the output signal are recorded.
Samples of such experimental oscillograms corresponding to the
points denoted by letters "a", "b", "c", "d" and "e" in
Fig.~\ref{fig:solifeldtheo15}  are presented in
Fig.~\ref{fig:oscillograms}. The left row of oscillograms
demonstrates that, when the external phase shift grows from
0.17$\pi$ to 0.95$\pi$ but the input power is constant (i.e.,
$S_0=0.75\pi$), the output signal profile with (a) two dark
solitons transforms into (b) three dark solitons, and then again
into (c) two dark solitons.

The right row of oscillograms in Fig.~\ref{fig:oscillograms}
demonstrates that with an increase of the input power at a fixed
value of the external phase shift 0.6$\pi$, the number of dark
solitons is increasing as the output oscillograms contain either
one, three, or five dark solitons. These results appear to be in a
good qualitative agreement with the theory~\cite{SKB}.

We should mention that the number of dark solitons observed in the
experimental profiles does not always match the number predicted
by the theory~\cite{SKB}. This is especially true for very small
and very large values of the input power. For small values of
$P_{\rm in}$, corresponding to the values $S_0 < 0.5\pi$, the dark
soliton is not properly formed as the time of the signal
propagation between the antennae is smaller than the so-called
"nonlinear time" $T_N = 1/(N |u|^2)$ during which the spin-wave
nonlinearity could significantly affect the pulse profile. On the
other hand, for large values of $P_{\rm in}$ corresponding to $S_0
> 1.5\pi$, nonlinear dissipation and other nonlinear effects, not
taken into account in our simple theoretical model, make the
dependence of $|u_0|^2$ on the input power nonlinear, which
prevents us from making a quantitative comparison between theory
and experiment. At the same time, it is clear from the
experimental data presented above that in the range of the
intermediate input powers, 0.5$\pi < S_0 <1.5\pi$, the theory
gives a good qualitative and even reasonable quantitative
explanation of the experimental data.

In conclusion, we have extended the concept of phase engineering,
demonstrated earlier for matter-wave dark solitons in
Bose-Einstein condensates, to spin-wave magnetic solitons. We have
demonstrated experimentally the crucial role played by an
externally introduced phase shift in the input pulse for the
process and outcome of the dark-soliton generation. Our
experimental results are in a good qualitative agreement with the
analytical and numerical predictions of dark-soliton generation,
and they provide more direct verification of the theory.

\section*{Acknowledgements}

We thank Carl Patton and Elena Ostrovskaya for useful discussions.
This project was supported by the Deutsche Forschungsgemeinschaft,
the Alexander von Humboldt Foundation, the MURI grant
W911NF-04-1-0247 of the US Army Research Office, and the
Australian Research Council.


\begin{thebibliography}{}
\bibitem{phasensprungdefinition} In Ref.~\cite{SKB} the definition of
the phase shift is slightly different. Here, only input pulses
with zero intensity in the dark part are considered, and
simplified phase jump was chosen in order to avoid confusion.
\end{thebibliography}
\end{document}